FRONT MATTER
**Title Atomically engineered cobaltite layers for robust ferromagnetism**
**Authors**
Shengru Chen,[1,2,†] Qinghua Zhang,[1,†] Xujing Li,[2,3,4,†] Jiali Zhao,[1] Shan Lin,[1,2] Qiao Jin,[1,2] Haitao Hong,[1,2] Amanda Huon,[5,6] Timothy Charlton,[5] Qian Li,[7] Wensheng Yan,[7] Jiaou Wang,[3] Chen Ge,[1] Can Wang,[1,2,8] Baotian Wang,[3,4] Michael R. Fitzsimmons,[5,9] Haizhong Guo,[10] Lin Gu, [1,2,8,‡] Wen Yin,[3,4,*] Kuijuan Jin,[1,2,8,*] Er Jia Guo[1,2,8,*]

**Affiliations**
[1] Beijing National Laboratory for Condensed Matter Physics and Institute of Physics, Chinese Academy of Sciences, Beijing 100190, China
[2] University of Chinese Academy of Sciences, Beijing 100049, China
[3] Institute of High Energy Physics, Chinese Academy of Sciences, Beijing 100049, China
[4] Spallation Neutron Source Science Center, Dongguan 523803, China
[5] Neutron Scattering Division, Oak Ridge National Laboratory, Oak Ridge, TN 37831, USA
[6] Department of Mathematics, Physics and Statistics, University of the Sciences, Philadelphia, Pennsylvania 19104, USA
[7] National Synchrotron Radiation Laboratory, University of Science and Technology of China, Hefei 230029, China
[8] Songshan Lake Materials Laboratory, Dongguan, Guangdong 523808, China
[9] Department of Physics and Astronomy, University of Tennessee, Knoxville, TN 37996, USA
[10] Key Laboratory of Material Physics, Ministry of Education, School of Physics and Microelectronics, Zhengzhou University, Zhengzhou 450001, China
[†]These authors contribute equally to the manuscript.
[‡] Present addresses: National Center for Electron Microscopy in Beijing, Tsinghua University, Beijing 100084, People's Republic of China, and School of Materials science and Engineering, Tsinghua University, Beijing, 100084, People's Republic of China
*Corresponding author. Emails: yinwen@ihep.ac.cn, kjjin@iphy.ac.cn, ejguo@iphy.ac.cn

**Abstract**
Emergent phenomena at heterointerfaces are directly associated with the bonding geometry of adjacent layers. Effective control of accessible parameters, such as the bond length and bonding angles, offers an elegant method to tailor competing energies of the electronic and magnetic ground states. In this study, we construct unit thick syntactic layers of cobaltites within a strongly tilted octahedral matrix *via* atomically precise synthesis. The octahedral tilt patterns of adjacent layers propagate into cobaltites, leading to a continuation of octahedral tilting while maintaining significant misfit tensile strain. These effects induce severe rumpling within an atomic plane of neighboring layers triggers the electronic reconstruction between the splitting orbitals. First-principles calculations reveal that the cobalt ions transits to a higher spin state level upon octahedral tilting, resulting in robust ferromagnetism in ultrathin cobaltites. This work demonstrates a design methodology for fine-tuning the lattice and spin degrees of freedom in correlated quantum heterostructures by exploiting epitaxial geometric engineering.
**Teaser** Unit-thick syntactic layers of cobaltites within a strongly tilted octahedral matrix exhibit a record-high ferromagnetism.

**MAIN TEXT**
**Introduction**

Heterointerfaces are connected by chemical bonds between dissimilar classes of materials (*1*). Corner-shared octahedral, along with metal–oxygen bonds, constitute the framework of oxide heterostructures (*2-4*). Conventionally, the most effective means to tune the lattice degree of freedom in a correlated material is to change the misfit strain or shear strain induced by substrates, buffer layers, capping layers, or crystal symmetry (*5-10*). Advances in the thin-film deposition method have enabled the atomically assembling of heterointerfaces by selectively fabricating oxide layers with different octahedral rotation patterns (*11, 12*). This phenomenon provides the octahedral rotation, which directly affects the bonding angle ($\beta_{B\text{-}O\text{-}B}$) and bond length ($r_{B\text{-}O}$), as an independent geometric constraint to tune the lattice structure. Interfacial control of the oxygen coordination environment allows an octahedra to rotate thereby retaining the octahedra of neighboring layers across interfaces and enabling functional properties. This methodology has been effectively applied to modify the magnetic anisotropy of SrRuO$_3$ (*13-16*) and La$_{2/3}$Sr$_{1/3}$MnO$_3$ (*17, 18*) thin layers. Similarly, the electronic states of transition metal ions also vary with structural modifications induced by octahedral rotations. The balance between crystal field energy ($\Delta_{\text{CF}} \propto r_{B-O}^{-5}$), intra-atomic exchange interaction energy ($\Delta_{\text{ex}}$), and bandwidth ($W \propto r_{B-O}^{-3.5} \cos(\pi - \beta_{B-O-B})$) not only substantially alters band structures but also affects the ground state of spin and orbital orders (*19, 20*). Thus, the octahedral rotation in oxide heterostructures offers a promising strategy to nanoengineer physical properties that are not observed in the bulk constituents or thick layers.

Rare earth scandates typically have heavily distorted orthorhombic perovskite structures. The single crystals of scandates are commonly used as substrates for epitaxial ferroelectric and ferromagnetic (FM) thin-film growth (*21-25*). The lattice of scandates is derived from the ideal cubic structure by tilting of ScO$_6$ octahedra along the [110]$_{\text{pc}}$ and [001]$_{\text{pc}}$ directions of pseudocubic unit cells (*26, 27*). For instance, the crystal structure of DyScO$_3$ (DSO) is shown in Fig. 1a. Dysprosium (Dy) atoms are surrounded by eight scandium–oxygen (ScO$_6$) octahedra. The octahedra are heavily tilted in response to the occupancy of *A*-site by a Dy cation, which is smaller than that required to form the ideal structure. The octahedral tilting reduces the crystalline symmetry to a *Pbnm* space group (*28, 29*). Theoretically, the tilting angle of ScO$_6$ octahedra in DSO, determined by crystal electrical field calculations, is up to ~25°(*30*), which is one of the largest octahedral tilting angles in perovskite oxides in the literature. In this work, we use this heavily tilted octahedral matrix in DSO to modify the octahedral tilt in the neighboring LaCoO$_3$ (LCO) layers, exhibiting an active spin-state transition upon structural distortions. We demonstrate that tensile strain induced by substrates and interlayer-mediated octahedral tilt is simultaneously present in synthetic atomic thin cobaltites. This approach successfully modifies electronic states by changing the bonding geometry. Robust large ferromagnetism is achieved in the single unit-cell LCO, which is non-magnetic in its original state. The propagation of octahedral tilt in long-range ordering suggests that this methodology can be readily applied to manipulate lattice–orbital interactions in other correlated heterogeneity materials.

DSO$_n$/LCO$_m$ (D$_n$L$_m$) superlattices (SLs) were grown on single crystalline (001)-oriented SrTiO$_3$ (STO) substrates by pulsed laser deposition, where *n* and *m* represent the numbers of unit cells (u. c.). The DSO/LCO bilayers in all SLs were repeated 10 times along the growth direction. X-ray diffraction (XRD) $\theta$–$2\theta$ scans demonstrate that all SLs are single phase and reciprocal space mapping results show that all SLs are epitaxially grown with high crystalline quality and coherently strained to STO substrates (Extended data Fig. S1–S3). The DSO ($a_o$ = 5.714 Å, $b_o$ = 5.438 Å, and $c_o$ = 7.897 Å) layers are [001]$_{pc}$-oriented (refers to the [110]$_o$ orientation in orthorhombic indices), where $o$ and $pc$ represent the orthorhombic and pseudocubic notation, respectively (*31*). The coherently grown DSO layers underwent a compressive strain of ~1.1% caused by the STO single crystal substrates, further increasing the octahedral tilt toward the in-plane direction (*32*). The real-space distorted lattice structure is visualized by atomically resolved scanning transmission electron microscopy (STEM). Figure 1b shows a cross-sectional annular bright field (ABF) STEM image of D$_1$L$_1$ SL. The STEM measurements were taken along the [110]$_{pc}$ zone axis, enabling the direct quantification of the $BO_6$ ($B$ = Sc, Co, and Ti) octahedral tilt. The single unit-cell LCO layers are sandwiched between two DSO layers. The octahedra of LCO are forced to follow the DSO's octahedral tilt character. The crystalline structure of simplified octahedra in DSO and LCO is schematically shown in Fig. 1b. In-phase versus out-of-phase octahedral tilts occur along the film growth direction. Simultaneously, the atomic plane within DSO or LCO experiences severe rumpling caused by octahedral tilts. The layer-position-dependent $\beta_{B-O-B}$ across the interfaces is summarized in Fig. 1b. In the interface regime, the $\beta_{B-O-B}$ decreases rapidly from $\beta_{Ti-O-Ti}$ ~ 180° in STO substrates. After the first five unit cells in SL, the octahedral tilt reduces to nearly a constant value $\beta_{B-O-B}$ ~140°. The coherency could be stabilized up to tens of bilayer repeats. Direct comparison of non-tilted and tilted octahedra in STO substrates and DSO/LCO SLs is shown in Figs. 1c and 1d. A high-angle annular dark field (HAADF) STEM image (Fig. 1e), along with electron energy loss spectroscopy (EELS) maps (Figs. 1f–1j), is recorded from a selected region in D$_1$L$_1$ SL. Atomic ordering in different chemical components confirms the designed growth sequence. The chemical intermixing across the interfaces is less than two atomic planes. SL exhibits ultrasharp interfaces as evidenced by HAADF STEM images and periodic Bragg peaks in XRD curves. These structural characterizations demonstrate that the non-ferromagnetic spacing layers act effectively as a template for engineering the octahedral tilting patterns of adjacent layers.

The magnetic properties of DSO/LCO SLs were characterized using a superconducting quantum interference device (SQUID) magnetometer. Field-cooled magnetization versus temperature ($M$-$T$) curves of SLs indicate a paramagnetic (PM)-to-FM phase transition at Curie temperature ($T_C$) ~ 75 K (Extended data Fig. S4). The lineshapes of $M$-$T$ curves for SLs are the same as those of LCO single layers, implying that the FM origin is from LCO ultrathin layers. We note that the $T_C$ of SLs is lower than that of an LCO thick film (*33*). This fact has been previously observed in other FM oxide films with reduced dimensionality. Field-dependent magnetization ($M$-$H$) curves for SLs are shown in Fig. 2a and Fig. S4 of Extended data. In contrast to the pure PM response from a DSO single layer (inset of Fig. 2a) (*34*), the $M$-$H$ curve of D$_1$L$_1$ SL exhibits an open hysteresis loop with non-zero remnant magnetization ($M_r$) of ~25 emu/cm$^3$ and coercive field ($H_C$) of ~650 Oe. This indicates that the long-range ferromagnetism persists in atomically thin LCO SLs. For a D$_{10}$L$_{10}$ SL, both $M_r$ and $H_C$ increase to ~32 emu/cm$^3$ and ~2 kOe, respectively. The increase in the number of magnetic parameters can be attributed to the thicker LCO layers, which can lead to the long-range spin ordering. Figures 2b and 2c show the saturation magnetization ($M_S$) and $H_C$ as a function of the averaged out-of-plane lattice constant of $D_nL_{10}$ SLs for $1 \leq n \leq 10$ (Extended data Figs. S3 and S4). The increase of out-of-plane lattice constants in SLs is caused by the increase of DSO layers' number in SLs. We observe large deviations in both curves of SLs as the lattice constant increases. The in-plane $H_C$ reaches the maximum value for SL ($m = 3$). The out-of-plane lattice constant of this SL is ~3.81 Å (*35*) which is equal to the pseudocubic lattice constant of bulk LCO. The observed anomaly in the magnetic properties indicates a remarkable structural modification effect.

In addition, although macroscopic magnetization measurements provide solid evidence on the ferromagnetism in atomically thin LCO, the pronounced PM signals from DSO at low temperatures prevent us from subtracting the exact $M_S$ of LCO. To separate the PM contribution from DSO, we performed polarized neutron reflectometry (PNR) measurements on a $D_{10}L_{10}$ SL to obtain magnetization across the layers of the sample. We choose the $D_{10}L_{10}$ SL as a test model not only because its total magnetization is similar to that of $D_1L_1$ SL, but also because the first Bragg peak appears within the neutron measurement range (*36-38*). The specular neutron reflectivities from spin-up ($R^+$) and spin-down ($R^-$) neutrons were collected as a function of wave vector transfer ($q$), as shown in Fig. 2d. The difference between $R^+$ and $R^-$ is further demonstrated by calculating the spin asymmetry [SA = $(R^+-R^-)/(R^++ R^-)$]. Non-zero SA confirms no evidence of antiferromagnetism within the SLs. PNR data were fitted using a model describing the chemical profile obtained from X-ray reflectivity fitting to the same SL (Extended data Fig. S5) (*39*). Open symbols and solid lines in Figs. 2d and 2e represent the experimental data and best fits, respectively. PNR fitting yields a reasonably small figure of merit value of ~0.05, indicating that the optimized chemical and magnetic models are applied. Figure 2f shows the depth profiles of nuclear and magnetic scattering length densities of $D_{10}L_{10}$ SL. The best fits show that the spins in LCO align parallel to that in DSO under applied magnetic fields. We calculate the in-plane magnetization of LCO and DSO layers within SLs. The first two bilayers exhibit comparably smaller magnetization than the rest in SLs. These variations are consistent with the observations of different octahedral tilts in the interfacial regime of SLs. From our analysis of the PNR results, we obtain $M_{DSO}$ = 122 ± 20 emu/cm$^3$ and $M_{LCO}$ = 43 ± 12 emu/cm$^3$, respectively, under a magnetic field of 0.5 T at 6 K. The obtained $M_{LCO}$ by PNR agrees well with our SQUID results in Fig. 2a.

Next, we analyze the impact of structural distortions on the electronic states by elemental specific X-ray absorption spectroscopy (XAS). Figure 3a shows XAS at O *K*- and Co *L*-edges for $D_1L_1$ SL. The reference XAS curves from an LCO tensile-strained film (*40*) and an HSrCoO$_{2.5}$ thin film (*41*) are present for direct comparison. XAS at Co *L*-edges consistently shows that the oxidization state of Co ions in SL is +3 with a negligible energy shift, suggesting no external effects, such as charge transfer across the heterointerfaces (*42, 43*) or any defects formation present in SLs (*44*). Additionally, the features at XAS O *K*-edges provide indirect but convenient means to reflect the unoccupied orbitals of transition metal ions in SLs. The distinct pre-peaks at ~530 eV correspond to the excitations from the O 1*s* state to the O 2*p*-Co 3*d* hybridized state (*45*) Thus, the XAS results suggest that the $t_{2g}$ bands in Co 3*d* orbitals are not fully occupied for $D_1L_1$ SL. In contrast, the absence of pre-peaks in HSrCoO$_{2.5}$ demonstrates that its $t_{2g}$ bands are completely occupied, similar to bulk LCO (*46*) Figure 4a shows the schematic of electronic structures and spin states as a response to the change in $\Delta_{CF}$ and $W$, i.e., $\beta_{B-O-B}$ and $r_{B-O}$. For the unstrained bulk LCO, degenerated orbitals are expected. Applying epitaxial strain would reduce orbital degeneracy and decrease the overlap between $d$ orbitals and neighboring oxygen orbitals, thereby reducing the energy of these orbitals. For tensile-strained LCO layers in $D_1L_1$ SL, the energy of $d_{x^2-y^2}$ orbital is lower than that of $d_{3z^2-r^2}$ orbital (Fig. 3b). Electrons excited from $t_{2g}$ bands preferentially occupy the $d_{x^2-y^2}$ orbital rather than the $d_{3z^2-r^2}$ orbital. The number of unoccupied states in the $d_{3z^2-r^2}$ orbital is larger than that in the $d_{x^2-y^2}$ orbital. The calculated X-ray linear dichroism (XLD) at Co $L$-edges confirms the expected anisotropic orbital occupancy in $e_g$ orbitals (Fig. 3c) (*47-49*) In $D_1L_1$ SL, $\beta_{Co-O-Co}$ decreases from a flattened bond, further reducing the $e_g$ bandwidth, which increases the number of occupied orbital states. The spin state of Co ions in $D_1L_1$ SL is presumably higher than that of thick LCO tensile-strained films with the flattened $\beta_{Co-O-Co}$. The evolution of orbital occupation states with octahedral tilt provides a reasonable explanation for the observed ferromagnetism in ultrathin LCO layers.

To validate our atomic engineering strategy on octahedral parameters, we grew another two sets of atomic thin LCO SLs using LaFeO3 (LFO) and SrTiO3 (STO) to replace DSO. The LFO single crystals have an orthorhombic structure with an identical $a^-a^-c^+$ tilt pattern in Glazer notation, as DSO (*50*). The lattice constant difference between DSO and LFO is only ~0.5%. Therefore, we expect a similar octahedral tilt in LFO from a high-symmetry cubic parent structure, accommodated by combined in-phase octahedral tilting about the orthorhombic *a*-axis and out-of-phase tilt about the orthorhombic *b*- and *c*-axes (Extended data Figs. S6 and S7). The octahedral tilting effectively and structurally modifies LCO layers. Fig. 4b shows the *M-H* hysteresis loop for an LFO1/LCO1 ($F_1L_1$) SL. Compared to the *M-H* loop for $D_1L_1$ SL, both SLs exhibit nearly identical $H_C$ under the same measuring conditions. We determined that both valence states of Fe and Co ions have +3, inhibiting interfacial charge transfer (Extended data Fig. S8). Presumably, the LFO layers maintain their antiferromagnetic ordering of the bulk, where five 3*d* electrons are arranged in a high spin (HS) state $t_{2g}^3 e_g^2$ and are coupled across the structures through super exchange interactions (*51*). The FM character of LCO was reinforced by measuring X-ray magnetic circular dichroism (XMCD) spectra for Co $L$-edges (Fig. 4c). Therefore, the magnetization of $F_1L_1$ SL is normalized to the total thickness of LCO layers. The $M_S$ of $F_1L_1$ SL reaches ~110 emu/cm$^3$ (~0.7 $\mu_B$/Co). This is the highest value of the net magnetic moment in atomic thin LCO layers reported to date. Another control experiment was conducted with a STO1/LCO1 ($S_1L_1$) SL, where STO has an $a^0a^0a^0$ non-tilted pattern (*17*). The octahedral tilt in LCO layers is strongly suppressed, as shown in the inset of Fig. 4b. The crystalline symmetry of LCO in $S_1L_1$ SL transits into a tetragonal-like structure. The spin state of Co$^{3+}$ ions would be lower than that in heavily tilted octahedra. Thus, the magnetization of $S_1L_1$ SL is negligible. The obvious discrepancy between the two cases (LCO/LFO vs. LCO/STO) provides strong evidence for octahedral distortion-mediated electronic states. Apparently, the intrinsic structural modification using the spacing layer is a different but effective means to influence the orbital occupancies. The consistent results of changing the active spin states of cobalt ions were reported to use the intrinsic crystal symmetry or the extrinsic hydrostatic pressure (*15, 52-54*).

To further clarify the effects of octahedral titling and lattice distortion on the spin state of LCO, we performed first-principles calculations within the framework of density functional theory (*55, 56*). Critical parameters, *e. g.,* the on-site Coulomb interaction ($U$) and exchange interaction ($J$), were chosen on the basis of the ground states of material systems (*57, 58*). The calculation details and selected critical parameters are summarized in Fig. S9 of Extended data. Both 2 × 2 × 2 and 3 × 3 × 3 supercells are constructed to simulate the structure of our SLs (Extended data Fig. S10). The structures were optimized with in-plane (biaxially tensile) lattice parameters of LCO fixed to that of STO substrates. The out-of-plane lattice constants of LCO were optimized by reducing the free energy of the geometric structure (Extended data Table S1). The initial lattice structure of LCO is nearly identical for both spin states (*59, 60*). Under tensile strain from STO substrates, the octahedral tilt in LCO is significantly enhanced simultaneously the octahedral tilt increases when LCO is proximate to DSO or LFO whereas the octahedral tilt in LCO is suppressed when it is sandwiched between two STO layers. We calculated the potential energies of the low spin configuration ($E_{LS}$) and intermediate spin configuration ($E_{IS}$) (Extended data Fig. S11). Figure 5a depicts the energy difference ($\Delta E = E_{IS} - E_{LS}$) between the two spin configurations as a function of octahedral rotation amplitude. $\Delta E$ is close to zero when the octahedral rotation amplitude is less than 2%. An increase in the rotation amplitude negates $\Delta E$, suggesting that electrons prefer to occupy HS states rather than LS states. The critical rotation amplitude for the spin-state transition reduces gradually when the LCO unit cell elongates along the growth direction. These results are consistent with our experimental findings that octahedral rotation enhances the macroscopic magnetization of atomic thin LCO layers. Furthermore, we calculated the average moment of DSO/LCO stacks as a function of the out-of-plane lattice constant, as shown in Fig. 5b. The magnetic moment of these stacks changes nonlinearly with structural elongation (increase of octahedral tilting angle), which agrees well with our experimental observations (Fig. 2c). The theoretical calculations reveal a strong correlation between octahedral rotation and electronic spin states in LCO, explaining the origin of macroscopic magnetization in atomic thin LCO layers.

**Discussion and conclusions**

In summary, we demonstrate the tuning of electronic states in atomic thin LCO by octahedral engineering. The octahedral tilt in LCO can be modified by neighboring layers and reaches a small Co–O–Co bonding angle of ~140°, which has not been reported previously. The heavily distorted octahedra trigger the active spin-state transition, resulting in a record-high magnetic moment in single unit-cell thick LCO layers. Effective manipulation of octahedral parameters in functional oxides enables the study of the impact of spin–orbital correlations in terms of magnetoelectric properties, paving a new way for controlling the magnetic functionalities that are not present in their bulk forms.

**Materials and Methods**

**Synthesis of superlattices**

The LCO-based SLs were grown on (001)-oriented STO substrates (Hefei Kejing Mater. Tech. Co. Ltd) using PLD. The STO substrates were pre-treated by HF acid and annealed at 1080 ºC for 90 mins to obtain atomically flat $TiO_2$-termination. The sintered LCO and LFO ceramics, and the STO and DSO single crystals were used as the ablation targets. The SLs were fabricated by alternating two targets for thin film growth. The thicknesses of individual layers are controlled by laser pulses. The substrate's temperature was kept at 750 ºC and oxygen partial pressure was maintained at 100 mTorr. After the thin film deposition, the samples were cooled down under the oxygen pressure of 100 Torr. The growth rate of each component was calibrated separately using XRR fittings.

**Structural and basic physical property characterizations**

XRD and RSM measurements were carried out using a Panalytical X'Pert3 MRD diffractometer. The macroscopic magnetizations of all SLs were measured by a SQUID magnetometer. All measurements were performed by applying in-plane magnetic fields. The *M-T* curves were recorded during the sample warm-up process after field-cooled at 1 kOe. The *M-H* hysteresis loops were recorded at 10 K. The *M-H* loops were obtained by subtracting the diamagnetic signals from STO layers and substrates. Unfortunately, the paramagnetic signals from DSO ultrathin layers cannot be subtracted because the paramagnetic contributions from DSO vary with layer thickness. Therefore, the magnetizations of DSO/LCO SLs were normalized to the total thickness of SLs. The LFO is an antiferromagnet and does not contribute to the measured magnetization, thus the magnetizations of LFO/LCO SLs were normalized to the total thickness of LCO layers.

**STEM measurements and analysis**

Cross-sectional TEM specimens of DSO/LCO and LFO/LCO SLs were prepared using $Ga^+$ ion milling after the mechanical thinning. HAADF and ABF imaging were carried out in scanning mode. The measurements were taken using JEM ARM 200CF microscopy at the Institute of Physics, Chinese Academy of Sciences. HAADF and ABF images were taken along the pseudocubic $[110]_{pc}$ zone axis in order to visualize the oxygen atoms and analysis the octahedral tilt in the lattices. The atomic positions of Co and O ions were determined by fitting the intensity peaks with Gaussian function. Therefore, the bond length and bonding angles of Co–O–Co (Sc–O–Sc) bonds can be obtained precisely. The obtained values were averaged over the single atomic plane within the ABF image. Error bars were extracted by calculating the standard deviation value. The elemental specific EELS mappings were performed by integrating the signals from a selected region after subtracting the exponent background using power law. All data were analyzed using Gatan Micrograph software.

**PNR measurements**

The PNR experiment on $D_{10}L_{10}$ SL was performed at the Magnetism Reflectometer (MR, BL-4A) of Spallation Neutron Source, ORNL. The sample was cooled down and measured at 6 K under an in-plane magnetic field of 0.5 T. PNR measurements were conducted in the specular reflection geometry with wave vector transfer ($q$) perpendicular to the surface plane. $q$ is calculated by $4\pi \sin(\alpha_i)/\lambda$, where $\alpha_i$ is the neutron incident angle and $\lambda$ is the wavelength of neutron beam. Neutron reflectivities from spin-up ($R^+$) and spin down ($R^-$) neutrons were recorded separately. To better illustrate their differences, the neutron reflectivities were normalized to the asymptotic value of the Fresnel reflectivity ($R_F = 16\pi^2/q^4$). Simultaneously, we calculate the spin asymmetry (SA) using $(R^+ - R^-)/(R^+ + R^-)$. By fitting the PNR data, we obtain the magnetizations of the DSO and LCO layers separately. Please note that we allow the magnetization of DSO layers to vary for each layer during PNR fitting. The standard deviations of the magnetization values of the layers form the uncertainties of the *M*s.

**XAS and XMCD measurements**

Elemental specific XAS measurements were performed on SLs at the beamline 4B9B of BSRF. All spectra were collected at room temperature in total electron yield (TEY) mode. The measurements were performed by changing the incident angle of linearly polarized x-ray beam. The sample's scattering plane was rotated by 30° and 90° with respect to the incoming photons. When the x-ray beam is perpendicular to the surface plane, the XAS signal directly reflects the $d_{x^2-y^2}$ orbital occupancy. While the angle between the x-ray beam and surface plane is 30°, the XAS signal contains orbital information from both $d_{x^2-y^2}$ and $d_{3z^2-r^2}$ orbitals. Therefore, for simplifying the results, the unoccupied in-plane orbital states are proportional to $I_{ip} = I_{90°}$, while the unoccupied out-of-plane orbital states can be calculated by $I_{oop} = (I_{90°} - I_{30°} \cdot \sin^2 30°)/\cos^2 30°$. XLD is calculated by $I_{ip} - I_{oop}$. The XMCD measurements were performed on $F_1L_1$ SL using circularly polarized x-rays at 77 K in TEY mode. Instead of switching the polarity of x-ray beams, we collected the XAS under the opposite magnetic fields of +/−0.15 T. The XMCD signals were obtained by calculating the difference between XAS (+0.15T) and XAS (−0.15 T).

**First-principles calculations**

The first-principles calculations were performed within the framework of density functional theory (DFT) with Perdew-Burke-Ernzerhof (PBE) exchange-correlation pseudopotential as implemented in the Vienna ab initio simulation package (VASP). A plane wave representation for the wave function with a cut off energy of 550 eV was applied. Geometry optimizations were performed using a conjugate gradient minimization until all the forces acting on the ions were less than 0.01 eV/Å per atom. A dense 14 × 14 × 14 Monkhorst-Pack K point grid was used for calculations. The crystal structures are built using VESTA software. The in-plane lattice parameters of superlattices, constrained by our x-ray diffraction measurements, were fixed to the STO substrate ($a$ = 3.905 Å). We use ISODISTORT to analyze the distorted modes and strain states of both experimental and optimized lattice structures. In all cases, we use the highly symmetrical *Pm-3m* structure to describe the crystal structure of LCO. The pseudo-cubic lattice constant of bulk LCO adapts $a_{pc}$ = 3.83 Å, which was obtained from earlier neutron diffraction data. For any applied lattice distortions, each formula unit maintains the same lattice volume same as the experiments. The rotation modes structure with a 2 × 2 × 2 supercell was adopted for the structural simulation. We normalized the rotation amplitude of 7% equals to the octahedral rotation angles $\beta_{Co-O-Co}$ of 154.33° with respect to its cubic structure (the ground states of lattice volume).

**Acknowledgements**

**Funding**: This work was supported by the National Key Basic Research Program of China (Grant Nos. 2020YFA0309100 and 2019YFA0308500), the National Natural Science Foundation of China (Grant Nos. 11974390, 11721404, 11874412, and 12174437), the Beijing Nova Program of Science and Technology (Grant No. Z191100001119112), the Beijing Natural Science Foundation (Grant No. 2202060), the Guangdong-Hong Kong-Macao Joint Laboratory for Neutron Scattering Science and Technology, and the Strategic Priority Research Program (B) of the Chinese Academy of Sciences (Grant No. XDB33030200). The XAS and XLD experiments were conducted at the beam line 4B9B of the Beijing

Synchrotron Radiation Facility (BSRF) of the Institute of High Energy Physics, Chinese Academy of Sciences. The XMCD experiments were performed at National Synchrotron Radiation Laboratory (NSRL) in Chia via user proposals. The PNR experiments were conducted at Magnetism Reflectometer (MR, BL-4A) at the Spallation Neutron Source (SNS), a DOE Office of Science User Facility operated by ORNL. **Author contributions:** These samples were grown and processed by S.R.C. under the guidance of E.J.G.; TEM lamellas were fabricated with FIB milling and TEM experiments were performed by Q.H.Z. and L.G.; theoretical calculations were performed by X.J.L., B.T.W., and W.Y.; XAS measurements were conducted by J.Z. and J.O.W.; XMCD measurements were conducted by Q.L. and W.S.Y.; PNR measurements were performed by A.H., T.C., and M.R.F.; S.R.C., S.L., Q.J., and H.T.H. worked on the structural and magnetic measurements. G.C., C.W., H.G. participated the discussions and provided important suggestions during the manuscript revision. E.J.G. initiated the research and supervised the work. S.R.C. and E.J.G. wrote the manuscript with inputs from all authors. **Competing interests:** The authors declare that they have no competing financial interests. **Data availability:** The data that support the findings of this study are available on the proper request from the first author (S.R.C.) and the corresponding authors (E.J.G., W.Y., and K.J.J.). **Additional information**: Supplementary information is available in the online version of the paper. Reprints and permissions information is available online. Certain commercial equipment is identified in this paper to foster understanding.

**Figures and figure captions**

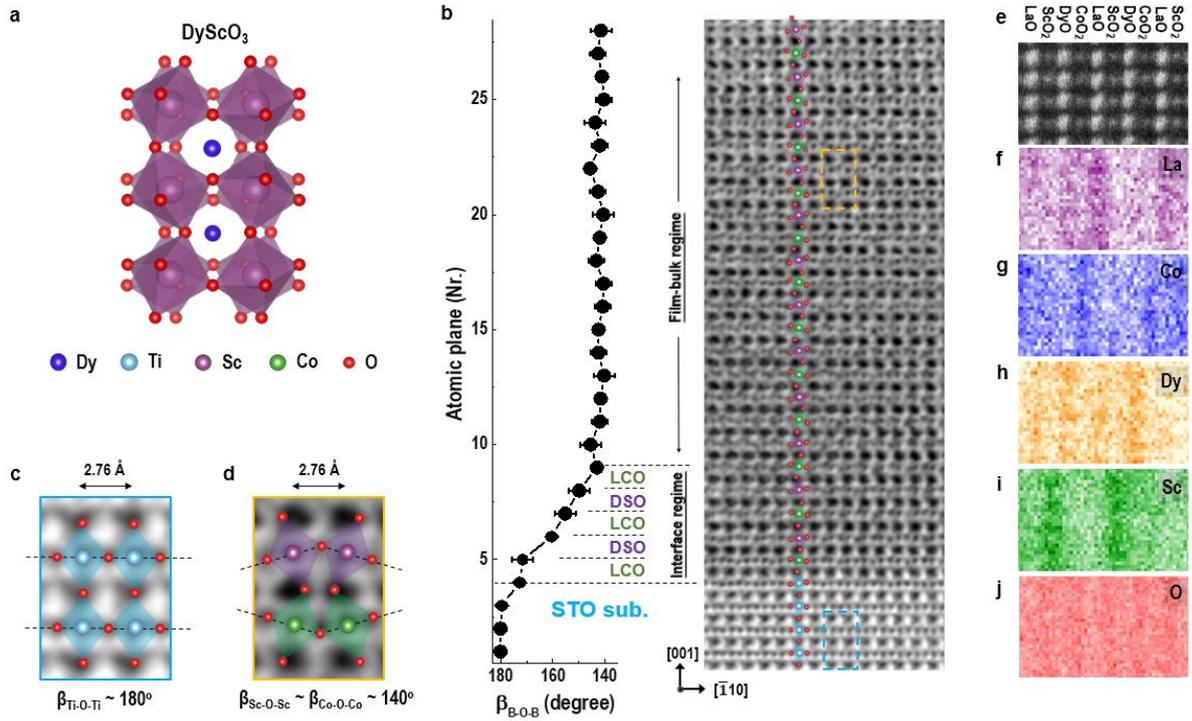

**Fig. 1. Structural characterizations of a single unit-cell LCO SL.** (a) Crystallographic structure of bulk $DyScO_3$, illustrating the heavily tilted octahedra within the framework. (b) Cross-sectional ABF STEM image of a $D_1L_1$ SL. The image was taken along the pseudocubic $[110]_{pc}$ zone axis. Layer position-dependent bonding angle ($\beta_{B-O-B}$) is shown on the left of the ABF image, where B represents the body-center cations (Co and Sc). $\beta_{B-O-B}$ was calculated by averaging bonding angles from atomic planes. Error bars represent one standard deviation. (c) and (d) Zoom-in representative oxygen octahedra in STO and LCO(DSO) layers marked in blue and yellow dashed squares in (b), respectively. Schematic illustrations indicate the orientations of octahedral tilts. (e) HAADF STEM image of a selected area from $D_1L_1$ SL. The colored panels show the integrated EELS intensities of (f) La $M$-, (g) Co $L$-, (h) Dy $M$-, (i) Sc $L$-, and (j) O $K$-edges, from the same region shown in (e).

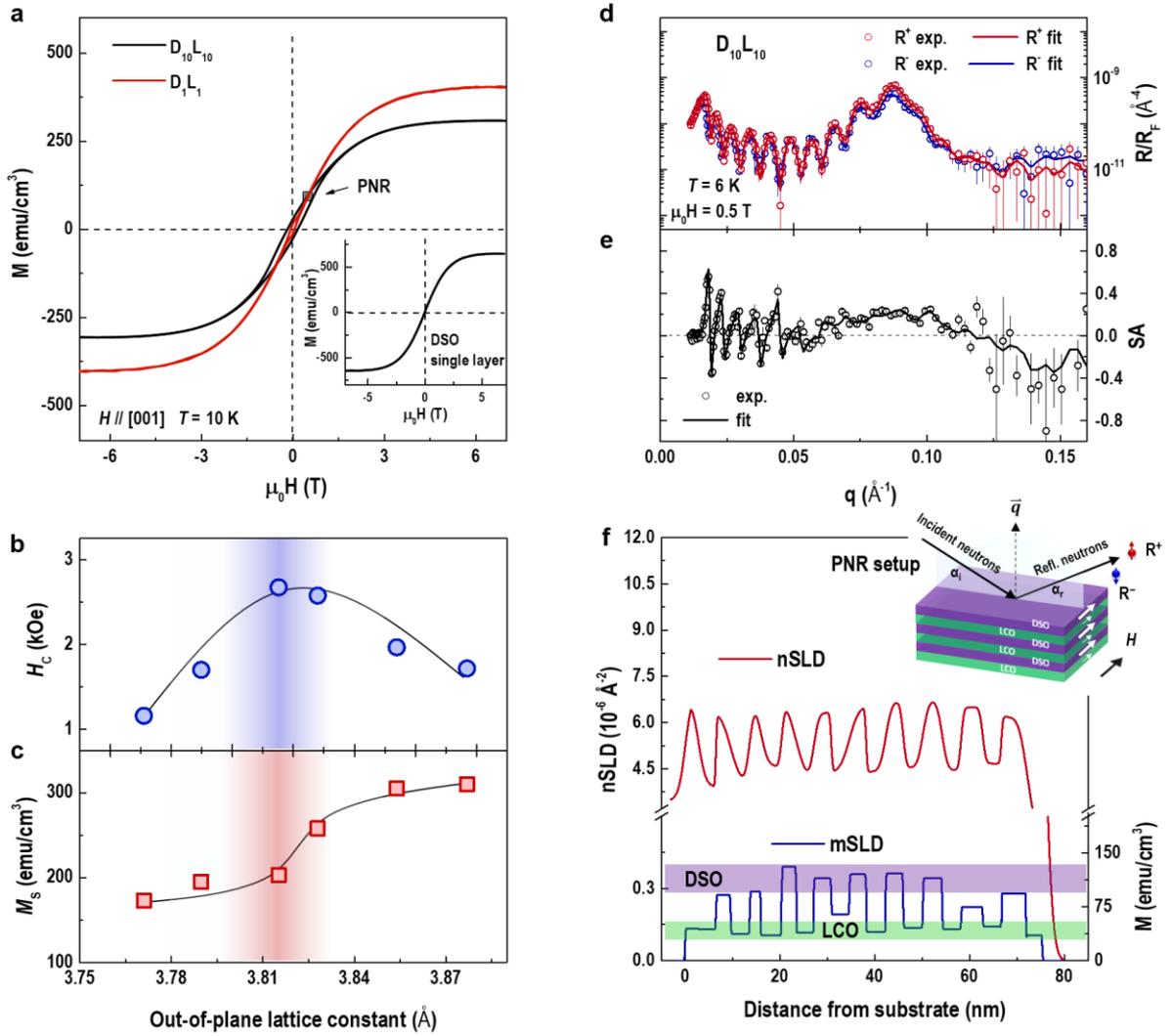

**Fig. 2. Magnetic properties of $D_nL_m$ SLs.** (a) *M-H* loops of $D_1L_1$ and $D_{10}L_{10}$ SLs. *M-H* loops were recorded at 10 K when a magnetic field was applied along the in-plane direction. Inset shows the *M-H* curve of a DSO single layer. (b) Coercive fields ($H_C$) and (c) saturation magnetization ($M_S$) of $D_nL_m$ SLs are plotted as a function of the out-of-plane lattice constant. $M_S$ was calculated using the total thickness of SLs. (d) Normalized neutron reflectivity curves and (e) spin asymmetry (SA) curve of a $D_{10}L_{10}$ SL for spin-up ($R^+$)- and spin-down ($R^-$)-polarized neutrons are shown as a function of the wave vector transfer *q*. (f) Nuclear scattering length density (nSLD) and magnetic scattering length density (mSLD) depth profiles of $D_{10}L_{10}$ SL. Magnetization of individual layers was calculated and compared with magnetometry data in (a). The inset of (f) displays the PNR measurement setup.

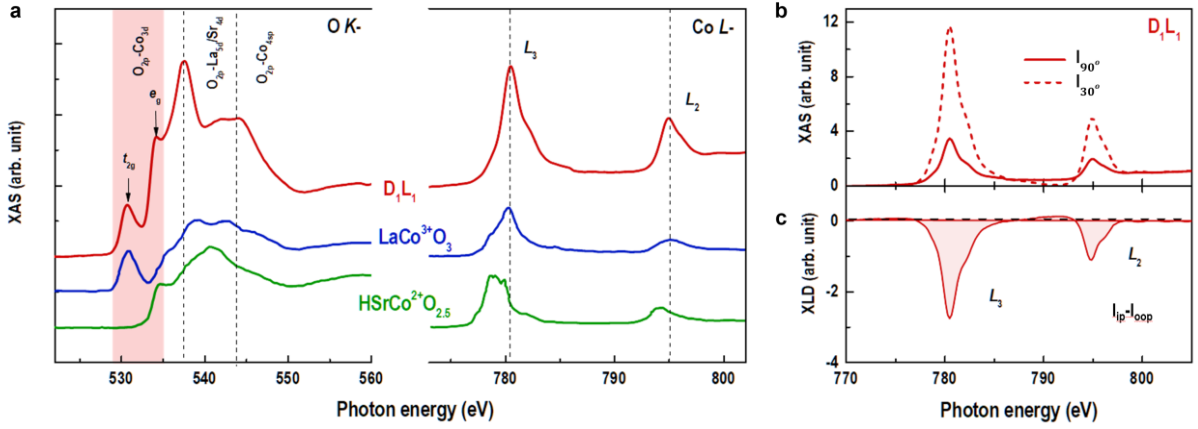

**Fig. 3. Increased $e_g$ orbital occupancy in single unit-cell LCO SL.** (a) Elemental specific XAS at O $K$- and Co $L$-edges for $D_1L_1$ SL. The shadow area in XAS O $K$-edges represents the electronic hybridization between O $2p$ and Co $3d$ orbitals. Reference data from an FM LaCo$^{3+}$O$_3$ tensile-strained thin film and an FM thin HSrCo$^{2+}$O$_{2.5}$ film are shown for direct comparison.[40, 41] (b) Polarization-dependent XAS at Co $L$-edges for $D_1L_1$ SL. The sample's scattering plane was rotated at angles of 90° and 30° with respect to the incident direction of an X-ray beam. Unoccupied in-plane orbital states are proportional to $I_{ip} = I_{90°}$, whereas unoccupied out-of-plane orbital states can be calculated by $I_{oop} = (I_{90°} - I_{30°} \cdot \sin^2 30°)/\cos^2 30°$. (c) XLD, calculated from ($I_{ip} - I_{oop}$), of $D_1L_1$ SL. The clear XLD signal demonstrates an asymmetric electronic occupancy in Co $3d$ $e_g$ orbitals.

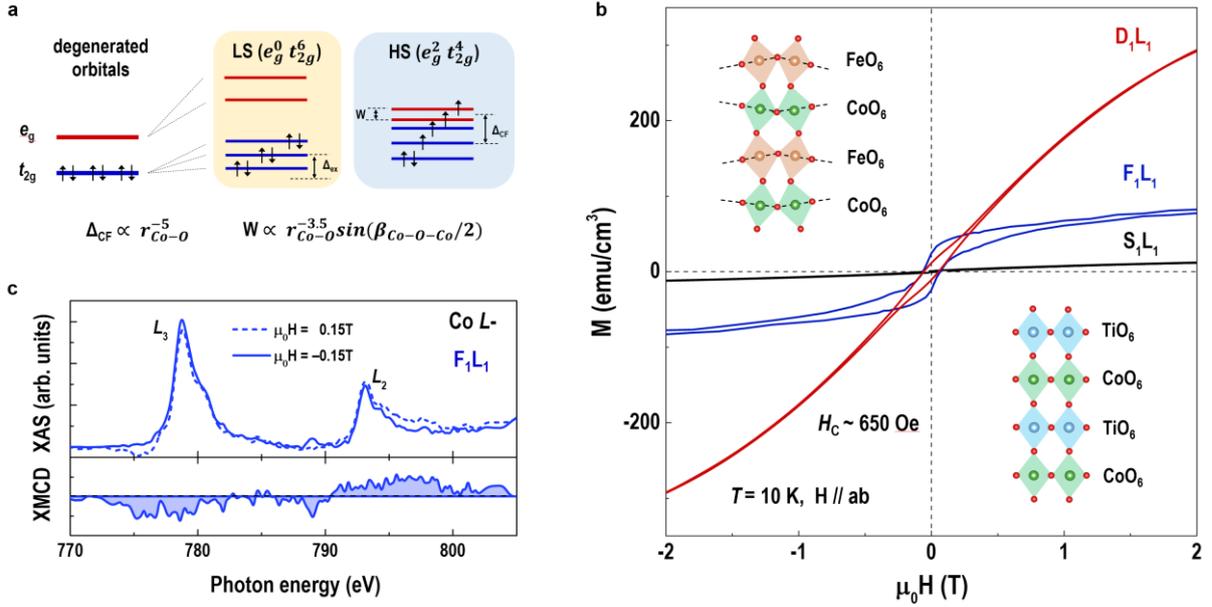

**Fig. 4. Manipulation of magnetic states in single unit-cell SLs via octahedral rotation**. (a) Electronic structure of Co 3$d$ orbital state. Degenerated $e_g$ and $t_{2g}$ orbitals split into two and three levels, respectively. Low spin (LS) and high spin (HS) states are effectively controlled by the interplay between $\Delta_{CF}$ and $W$. (b) $M$-$H$ loops of $D_1L_1$, $F_1L_1$, and $S_1L_1$ SLs. Magnetization measurements were performed at 10 K and in an in-plane magnetic field. Both $D_1L_1$ and $F_1L_1$ SLs exhibit identical $H_C$ ~650 Oe. Insets show octahedral rotation patterns in $D_1L_1$ and $F_1L_1$ SLs. (c) XAS and XMCD at Co $L$-edges for a $F_1L_1$ SL. XAS spectra were collected by applying a circularly polarized X-ray beam at 77 K under magnetic fields of +0.15 T (dashed line) and −0.15 T (solid line). XMCD spectra were calculated from the difference between two XAS spectra, demonstrating a clear magnetic signal from Co ions.

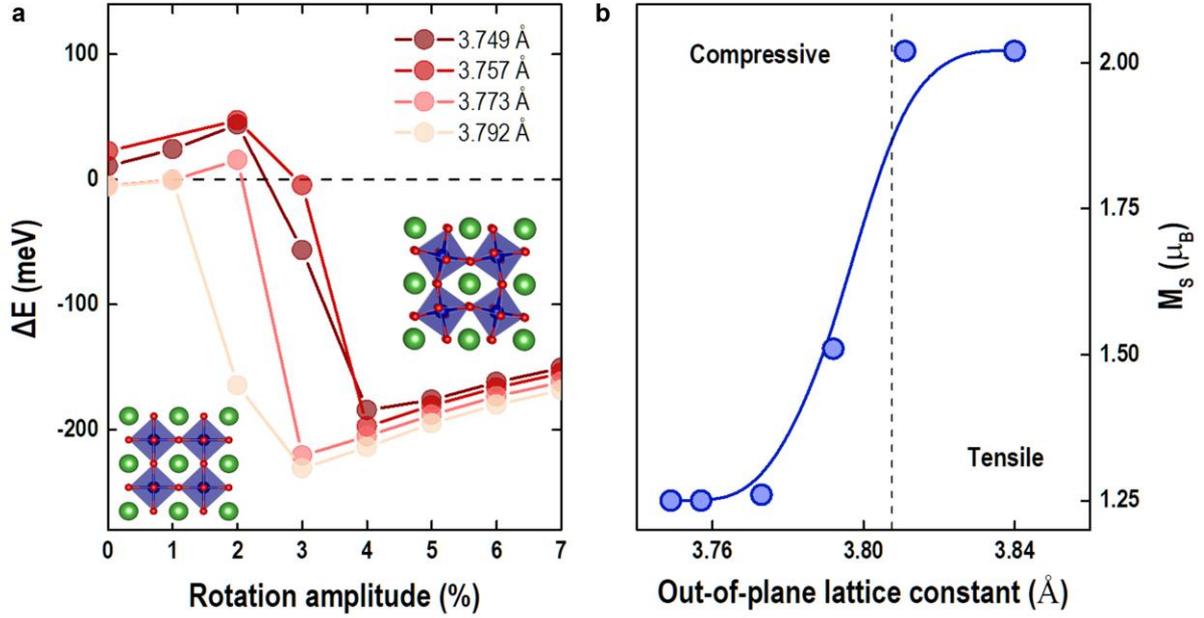

**Fig. 5. First-principles calculations on the electronic states of SLs at different rotation angles and under different lattice distortions.** (a) The energy difference ($\Delta E = E_{IS} - E_{LS}$) between intermediate spin (IS) and low spin (LS) states as a function of the octahedral rotation amplitude. The rotation amplitude varying from 0% to 7% represents $\beta_{Co\text{-}O\text{-}Co}$ changing from 180° to 153°, respectively. Insets show the schematic crystal structures of 2 × 2 LCO atomic layers with and without octahedral tilt. (b) Nonlinear change of magnetic moments in LCO with out-of-plane lattice constants. Lattice deformation is constrained by ISODISTORT.

# Supplementary Materials for

**Atomically engineered cobaltite layers for robust ferromagnetism**

Shengru Chen, Qinghua Zhang, Xujing Li, Jiali Zhao, Shan Lin, Qiao Jin, Haitao Hong, Amanda Huon, Timothy Charlton, Qian Li, Wensheng Yan, Jiaou Wang, Chen Ge, Can Wang, Baotian Wang, Michael R. Fitzsimmons, Haizhong Guo, Lin Gu, Wen Yin,* Kuijuan Jin,* Er Jia Guo*

*Corresponding author. Email: yinwen@ihep.ac.cn, kjjin@iphy.ac.cn, ejguo@iphy.ac.cn

This PDF file includes:

**Supplementary Materials**

    Figure S1. Structural characterizations of $D_1L_1$ SL.
    Figure S2. Structural characterizations of $D_{10}L_m$ SLs.
    Figure S3. Structural characterizations of $D_nL_{10}$ SLs.
    Figure S4. Magnetization characterizations of $D_nL_{10}$ SLs.
    Figure S5. XRR of $D_{10}L_{10}$ SL.
    Figure S6. Structural characterizations of $F_1L_1$ SL.
    Figure S7. STEM results of $F_1L_1$ SL.
    Figure S8. XAS measurements of $F_1L_1$ SL.
    Figure S9. Structural parameters and calculation details.
    Figure S10. Crystal structures with different tilting patterns.
    Figure S11. Projected density of states (DOS) for different spin states LaCoO3 with rotation amplitude varying from 0% to 7%.
    Table S1. Structural parameters for an initial LCO with a rotation pattern between $a^-a^-a^-$ and $a^0a^0a^0$.
    References (R1 to R6)

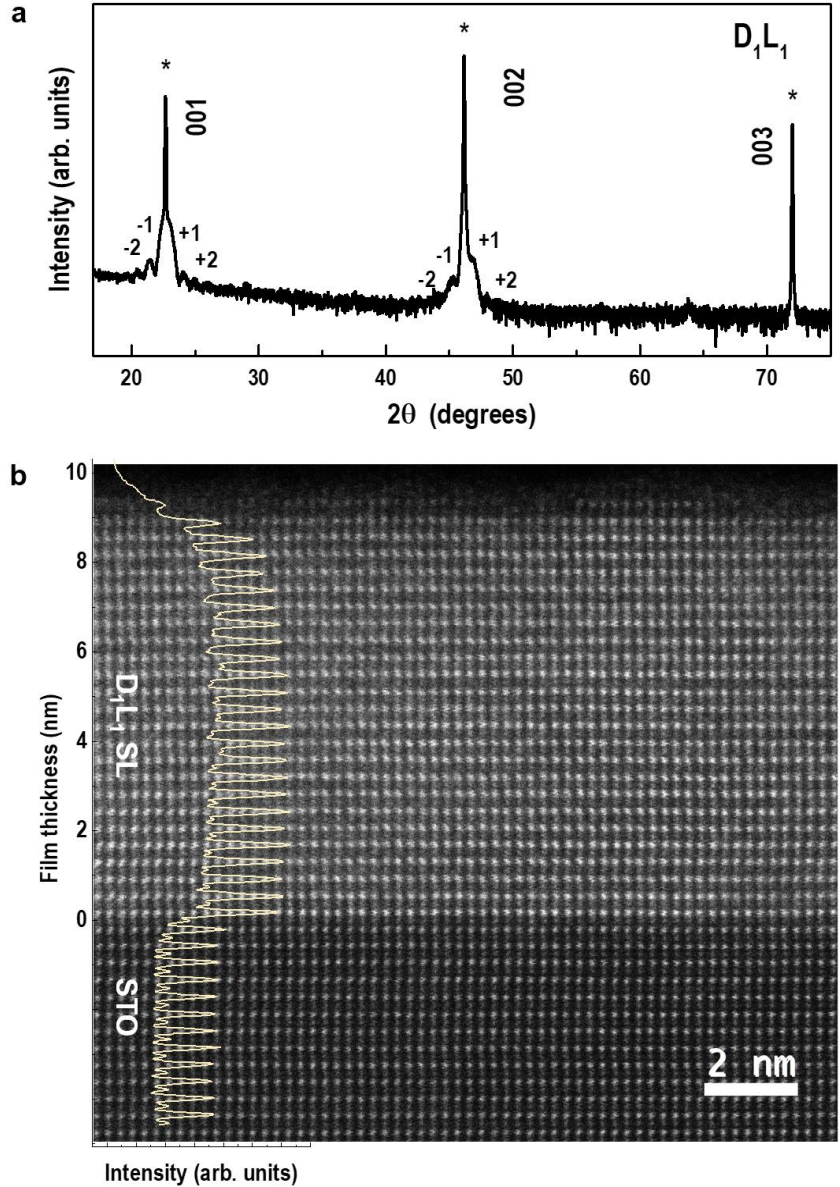

**Figure S1. Structural characterizations of $D_1L_1$ SL.** (a) XRD $\theta$-$2\theta$ scan of $D_1L_1$ SL. The "*" denotes the STO substrate's $00l$ reflections. The $00l$ reflections from $D_1L_1$ SL are overlapped with substrate's reflections. The clear Kiessig fringes around film's peaks suggest that SL is epitaxially grown with extremely high quality. (b) Cross-sectional HAADF STEM image of $D_1L_1$ SL. The sample was imaged along the pseudocubic [110] zone axis. Inset shows the HAADF intensity as a function of film thickness. The heavier elements (Sr, La, Dy) with the larger atomic number show brighter features in HAADF image. From STEM image and depth profile, we find the distinct interfaces between SLs and STO substrates are atomically sharp.

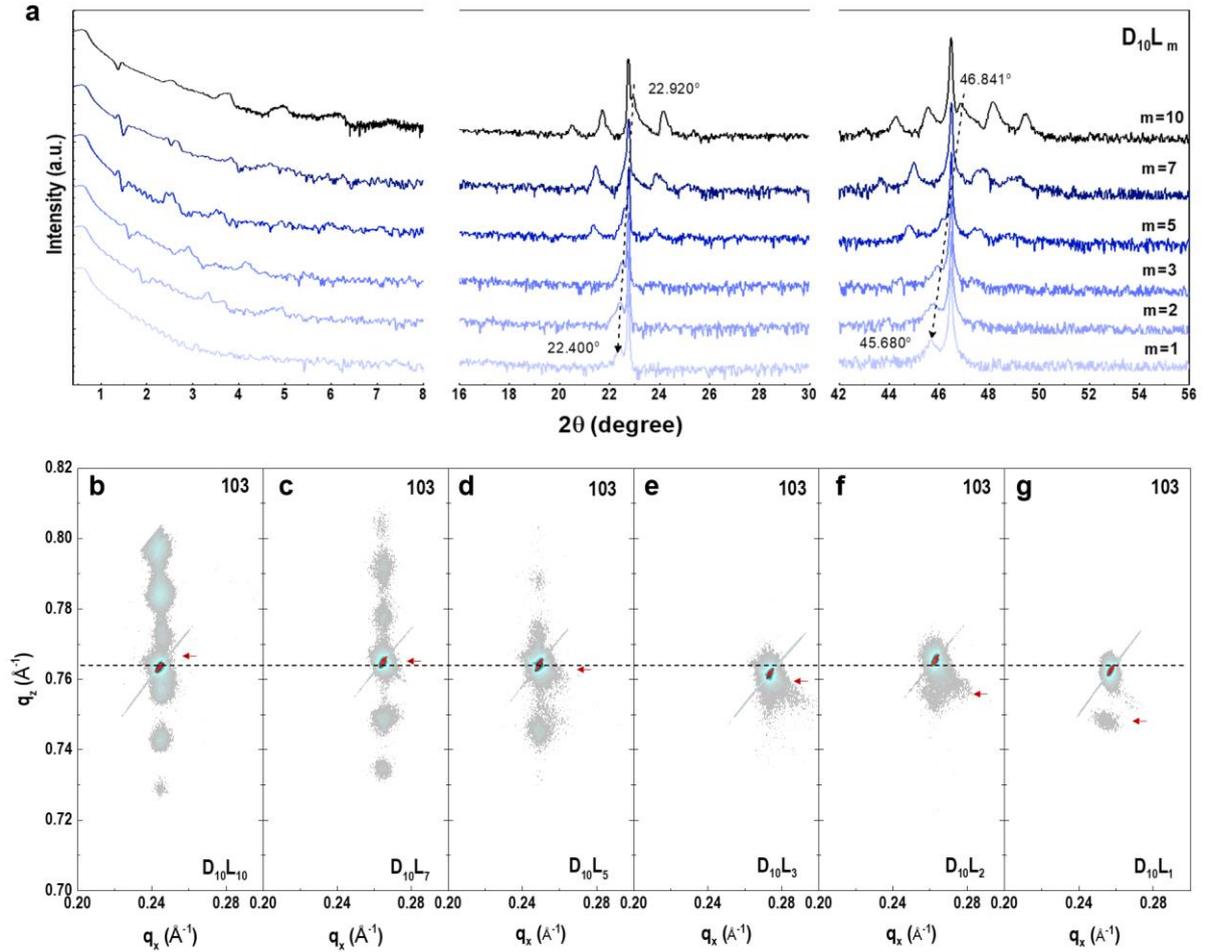

**Figure S2. Structural characterizations of $D_{10}L_m$ SLs.** (a) XRD $\theta$-$2\theta$ scans and (b) Reciprocal space mappings (RSMs) around the substrate's 103 reflections of $D_{10}L_m$ SLs for $m$ =1, 2, 3, 5, 7, 10, where $m$ represents the number of LCO's unit cell (u. c.). As increasing $m$, the SL peak shifts towards the large angles, suggesting the averaged out-of-plane lattice constant of SL reduces. The Kiessig fringes around the SL's main peaks and Bragg peaks indicate that all SLs are highly epitaxially grown and have the high crystalline quality. RSM results suggest all SLs are coherently strained by substrates. Red arrows indicate that the peak positions of SLs gradually shift to smaller $q_z$, indicating the increment of lattice constant.

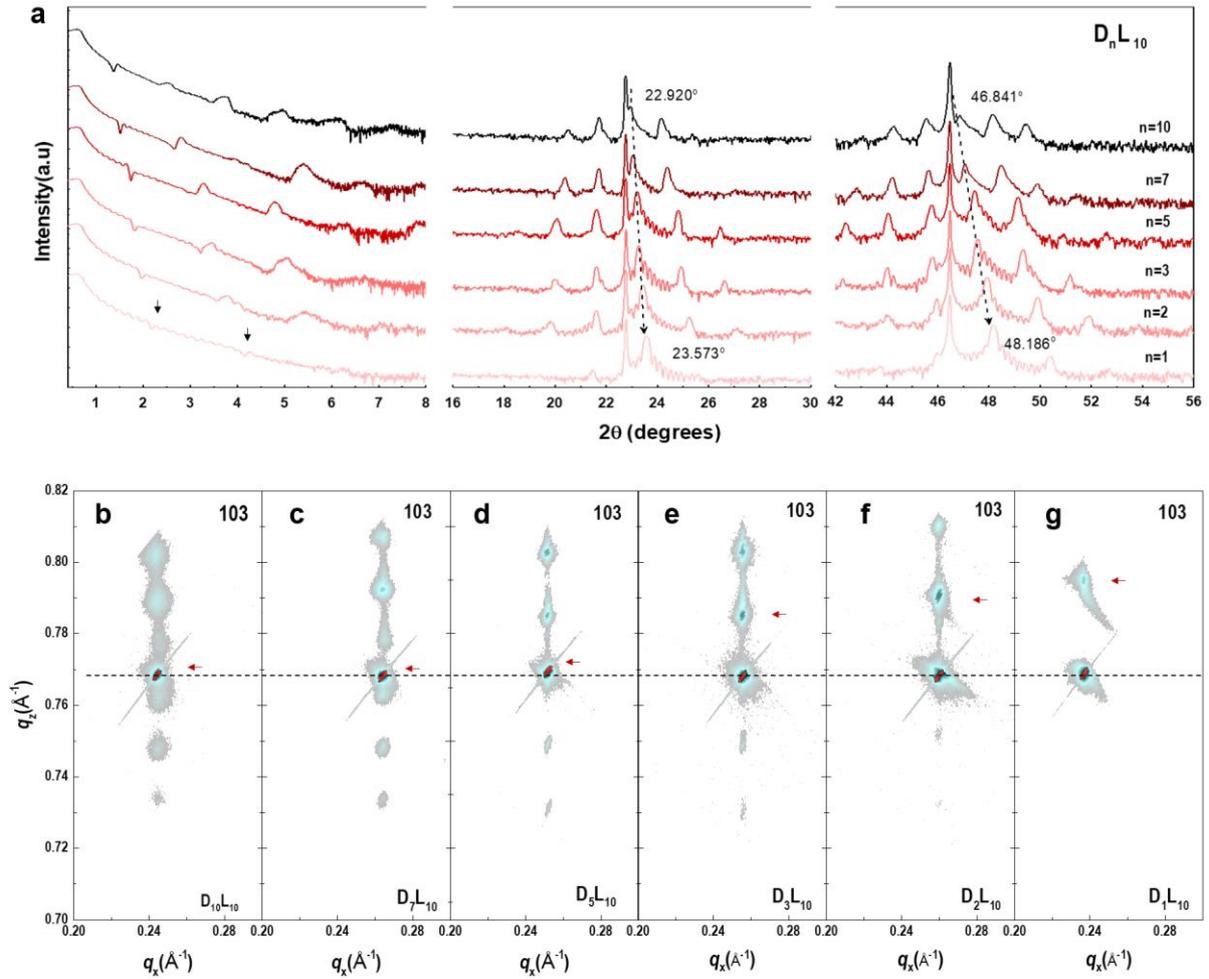

**Figure S3. Structural characterizations of $D_nL_{10}$ SLs.** (a) XRD $\theta$-$2\theta$ scans and (b) RSMs around substrate's 103 reflections of $D_nL_{10}$ SLs for $n$ =1, 2, 3, 5, 7, 10, where n represents the number of DSO's u. c.. With increasing DSO layer thickness, the SL peaks move to low angles, indicating the averaged out-of-plane lattice constant of SL increases. We observe up to 3 orders of SL Bragg peaks and thickness fringes around SL main peaks, suggesting that all SLs have high crystallinity. Similar to Figure S2, RSM results suggest all SLs are coherently strained by substrates.

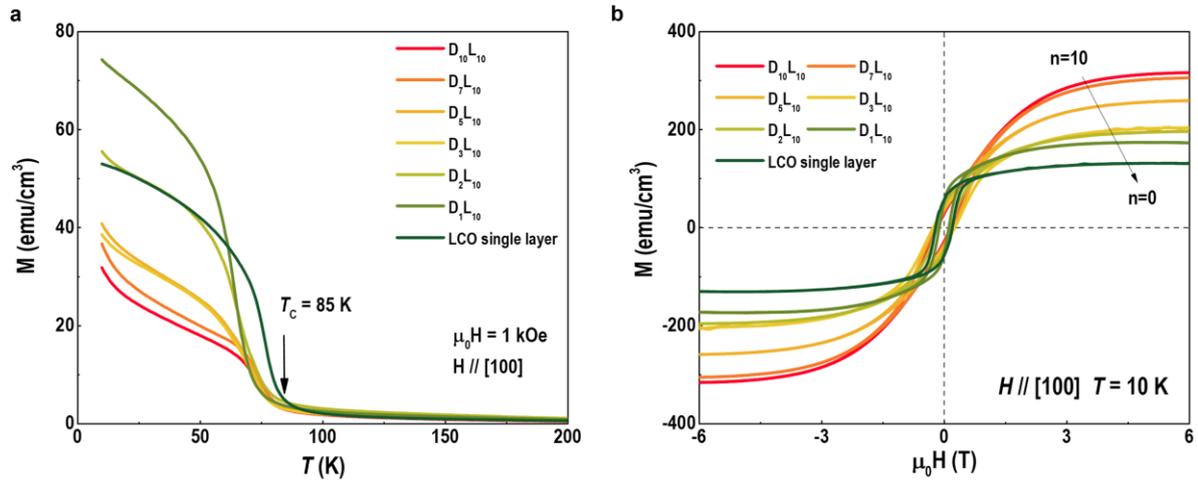

**Figure S4. Magnetization characterizations of $D_nL_{10}$ SLs.** (a) *M-T* curves of $D_nL_{10}$ SLs for $1 \leq n \leq 10$. The *M-T* curves were measured under a magnetic field of 1 kOe applied along the [100] orientation after field-cooling. *M-T* results show that LCO single layer and all SLs exhibit clear paramagnetic-ferromagnetic phase transitions. $T_C$ of LCO single layer is ~ 85 K, [R1] whereas $T_C$ of SLs decreases to 75 ± 3 K. The reduction of $T_C$ is attributed to the finite size effect. [R2] (b) *M-H* loops of $D_nL_{10}$ SLs. *M-H* loops were recorded at 10 K under in-plane magnetic field in parallel to [100] orientation. All samples exhibit clear hysteresis loops, indicating the ferromagnetic character at low temperatures. As increasing DSO layer's thickness, the saturation moment increases due to the paramagnetic contribution from DSO layers. [R3-R5]

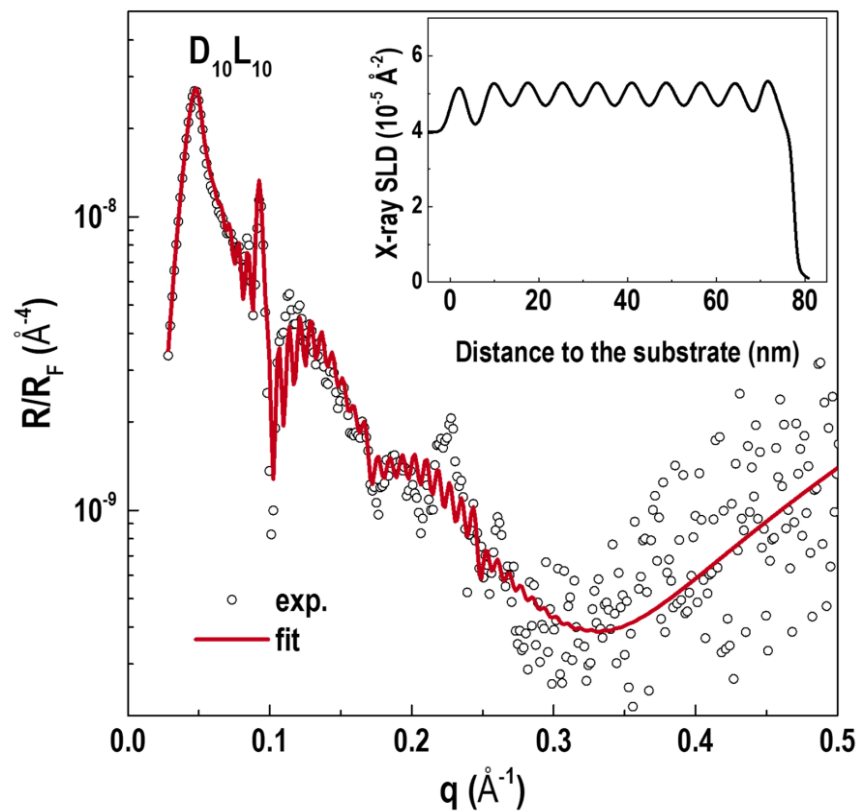

**Figure S5. XRR of $D_{10}L_{10}$ SL.** The solid line is the best fit to the experimental data (open symbols). The thicknesses of DSO and LCO layers are 39.2 ± 4.6 Å and 38.5 ± 3.9 Å, respectively. The DSO/LCO bilayer repeats 10 times. The total thickness of $D_{10}L_{10}$ SL is 79.8 ± 0.9 nm. Inset shows the X-ray scattering length density (SLD) profile of $D_{10}L_{10}$ SL. The X-ray SLD of LCO layer (~5.2×10$^{-5}$ Å$^{-2}$) is larger than that of DSO layer (~ 4.5×10$^{-5}$ Å$^{-2}$). The chemical composition of SL is used to constrain the chemical depth profile for PNR fittings in Figs. 2d-2f. We use the GenX software to fit XRR and PNR curves. [R6]

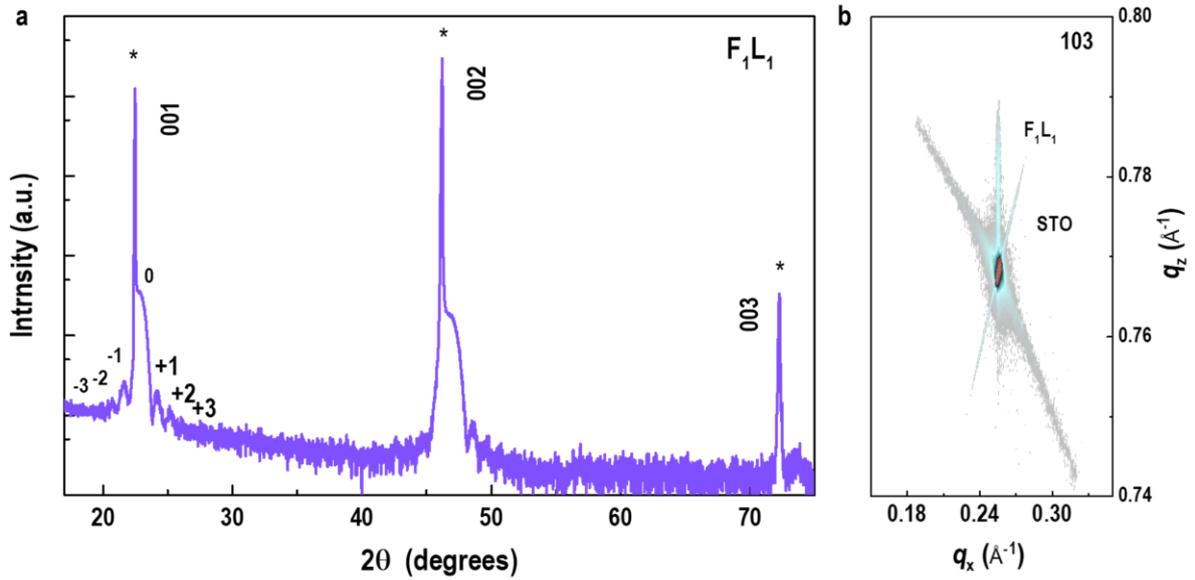

**Figure S6**. **Structural characterizations of $F_1L_1$ SL**. (a) XRD $\theta$-$2\theta$ scan of $F_1L_1$ SL. The "*" denotes the STO substrate's $00l$ reflections. The lattice constant of LFO is smaller than that of DSO, thus the averaged lattice constant of $F_1L_1$ SL is smaller than that of $D_1L_1$ SL. We observe the main peak of SL shifts to the right side of the substrate's reflections. Similar to Figure S1, the clear Kiessig fringes around the 001 and 002 peaks persist up to the 3$^{rd}$ order, indicating that SL is epitaxially grown with extremely high crystalline quality. (b) RSM around substrate's 103 reflections of $F_1L_1$ SL, suggesting all layers are coherently strained to STO substrates.

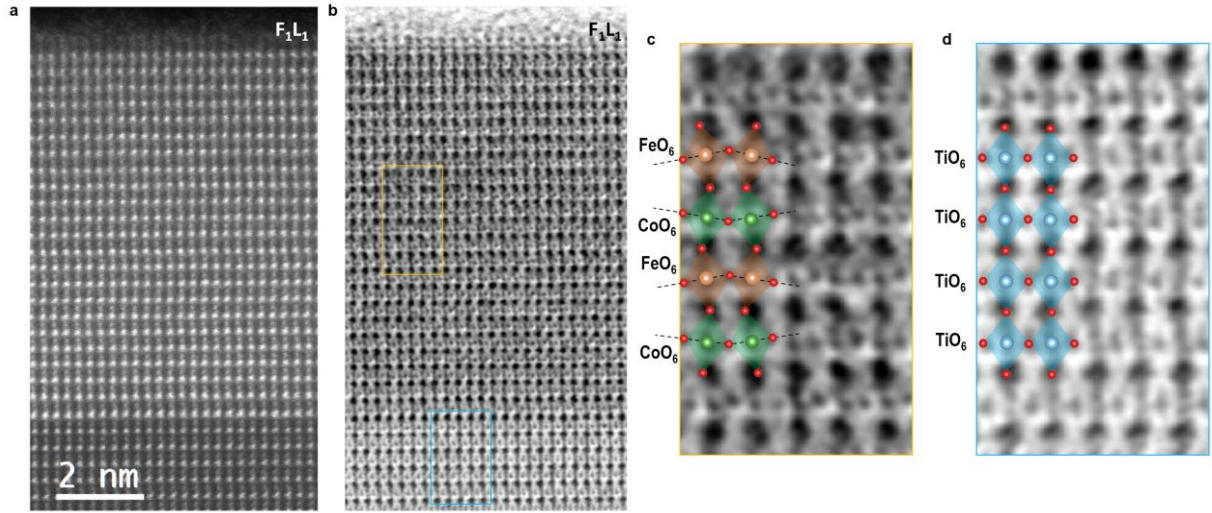

**Figure S7. STEM results of $F_1L_1$ SL**. Cross-sectional (a) HAADF-STEM and (b) ABF-STEM images of a $F_1L_1$ SL. The number of atomic layers is clearly visualized to confirm the designed structure. The interfaces between SL and STO substrates are proven to be atomically sharp. STEM results manifest itself a coherent and alternative LFO-LCO layers with a fully-strained state. The large tensile strain up to ~2.5% is applied to LCO layers. Zoom-in ABF images marked in orange and blue rectangles in (b) represent SL layers and STO substrates, as shown in (c) and (d), respectively. In contrast to untilted $TiO_6$ octahedra in STO substrates, the $CoO_6$ octahedra follow the tilt patterns of $FeO_6$ octahedra. We could identify the octahedral tilt angle is ~15° ± 3°, corresponding to $\beta_{Co-O-Co}$ ~150° ± 5°. This feature is similar to the octahedral tilt in $D_1L_1$ SL, as shown in Figure 1. STEM results highlight the importance of octahedral tilting in controlling the spin states of transition metal ions.

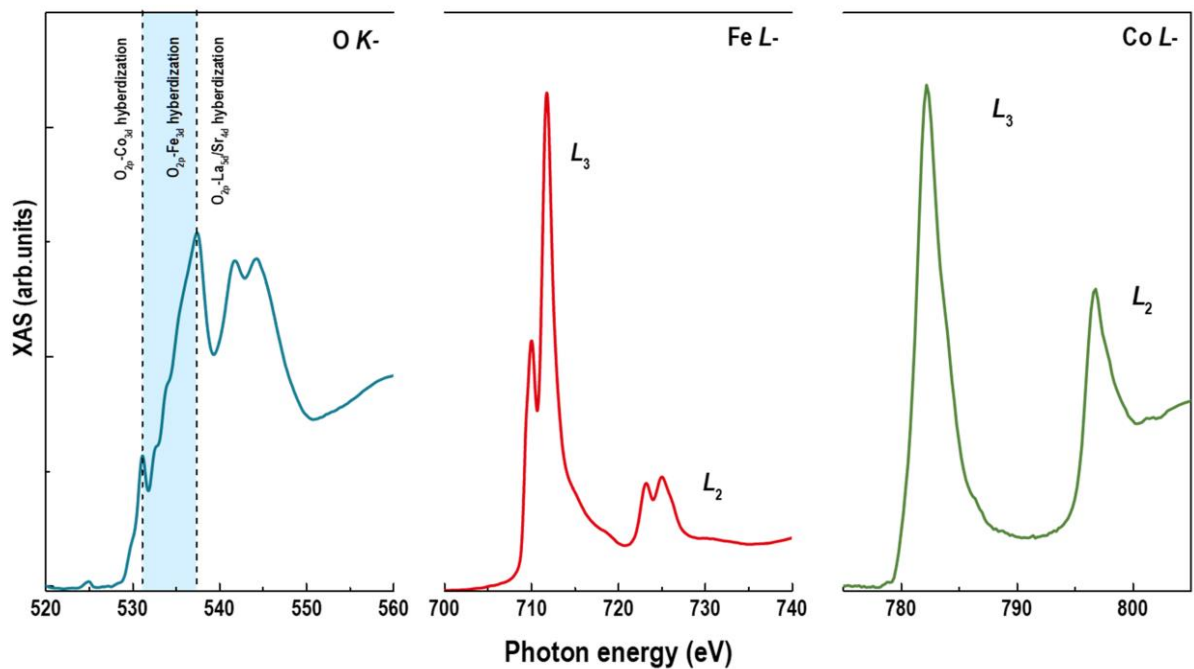

**Figure S8. XAS measurements of $F_1L_1$ SL**. XAS at O $K$-, Fe $L$-, and Co $L$-edges were measured at room temperature. XAS results indicate that Fe ions and Co ions keep +3, indicating the negligible charge transfer between Co and Fe ions. This fact is important for analyzing the spin states of Co ions within $F_1L_1$ SL.

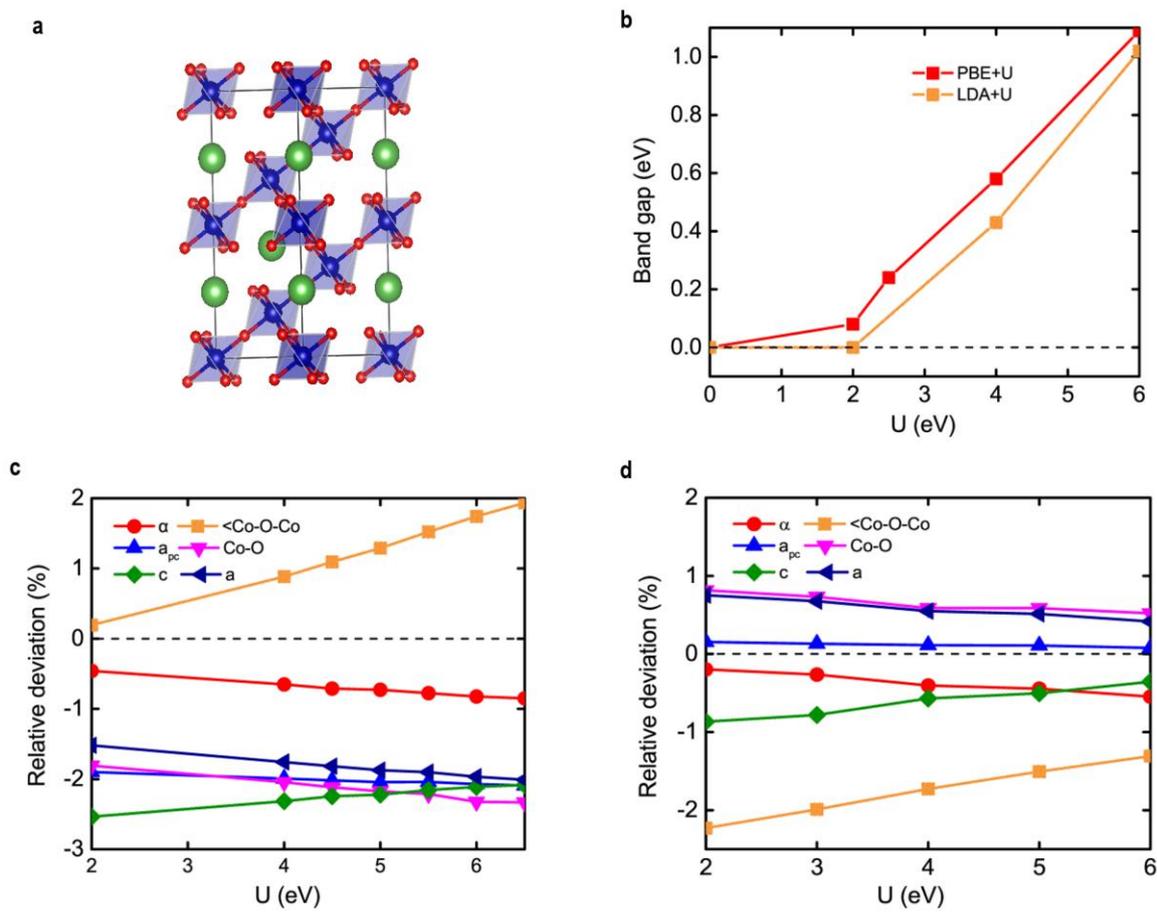

**Figure S9. Structural parameters and calculation details.** (a) The crystal structure of bulk LCO. (b) Band gaps of the optimized low spin state LCO using LSDA+U and sPBE+U method. (c) and (d) Lattice constants (*a,c*), pseudo-cubic constants ($a_{pc}$), Co-O-Co bond lengths, rhombohedral tilting angles (α), and Co-O-Co bond angles for the optimized low spin state LCO using LDA+U and PBE+U methods, respectively. Both methods exhibit similar trends, verifying the intrinsic property of LCO with different on-site Coulomb interactions.

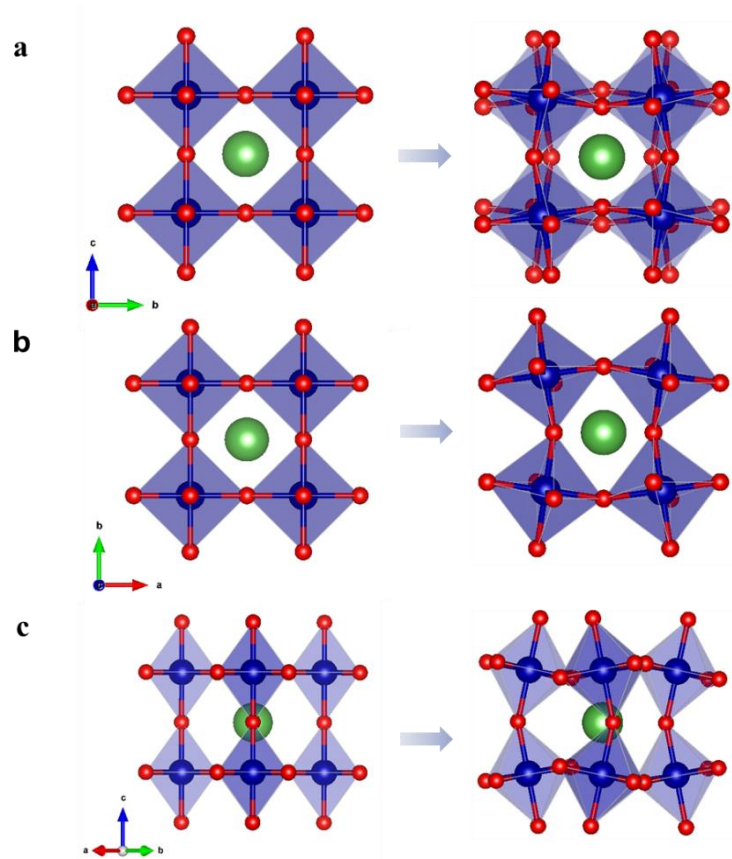

**Figure S10. Crystal structures with different tilting patterns.** (a), (b), and (c) Schematic crystal structures of 2 × 2 × 2 supercells along the pseudocubic [100], [001], and [110] zone axis, respectively. The supercells are without (lefthand) and with (righthand) octahedral tilt, corresponding to the $a^0a^0a^0$ non-tilted pattern ($S_1L_1$ SL) and $a^-a^-c^+$ tilt pattern ($D_1S_1$ and $F_1L_1$ SLs), respectively.

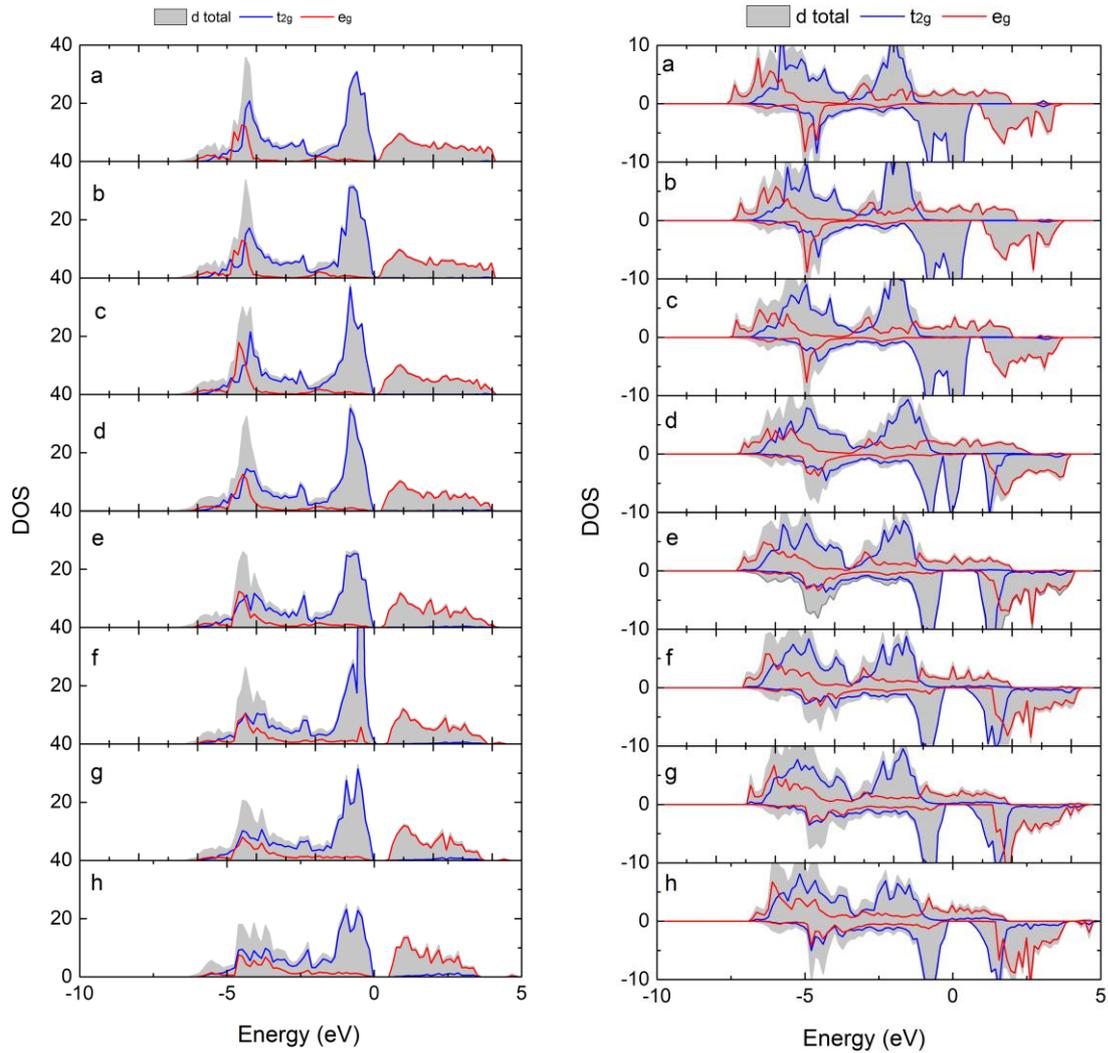

**Figure S11. Projected density of states (DOS) for LS (lefthand) and IS (righthand) states LaCoO$_3$ with rotation amplitude varying from 0% to 7%.** The gray area, bule lines, and red lines represent the Co ions $d$ total, $t_{2g}$ and $e_g$, respectively.

**Table S1. Structural parameters for an initial LCO with a rotation pattern between $a^-a^-a^-$ and $a^0a^0a^0$.** The lattice parameters were optimized using DFT calculations. Both tilted and non-tilted lattice structures are considered. The in-plane lattice parameter (*a*) of LCO is constrained by that of STO substrates. The out-of-plane lattice parameter (*c*) of LCO is optimized with minimizing the free energy. $\beta_{B-O-B}$ and $\beta'_{B-O-B}$ represent the octahedral rotation angle clockwise or counter-clockwise with respect to the in-plane direction. We summarize the calculated structural parameters of LCO with LS (IS) configuration, as listed below.

|  | strain (%) | a (Å) | c (Å) | $\beta_{B-O-B}$ (°) | $\beta'_{B-O-B}$ (°) |
|---|---|---|---|---|---|
| **Without tilting** | 2 | 3.905 | 3.77(3.81) | 180(180) | 180(180) |
| **With tilting** | 2 | 3.905 | 3.78(3.85) | 161.21(160.35) | 155.85(159.7) |

**Supporting Information References**